\documentclass[letterpaper]{article}
\usepackage{iccc}

\usepackage[pdftex]{graphicx}
\graphicspath{{./figures/}}
\DeclareGraphicsExtensions{.pdf,.jpeg,.png}

\usepackage{times}
\usepackage{helvet}
\usepackage{courier}

\usepackage[dvipsnames]{xcolor}
\usepackage{url}

\title{Calliope: An Online Generative Music System for Symbolic Multi-Track Composition}
\author{Renaud Bougueng Tchemeube \\
rbouguen@sfu.ca\\
\And
Jeff Ens\\
jeff\_ens@sfu.ca\\
School of Interactive Arts and Technology\\ 
Simon Fraser University\\
Surrey, BC V3T 0A3, Canada\\
\And
Philippe Pasquier\\
pasquier@sfu.ca\\
}

\begin{document} 
\maketitle

\begin{figure*}
	\centering
	\includegraphics[width=\linewidth]{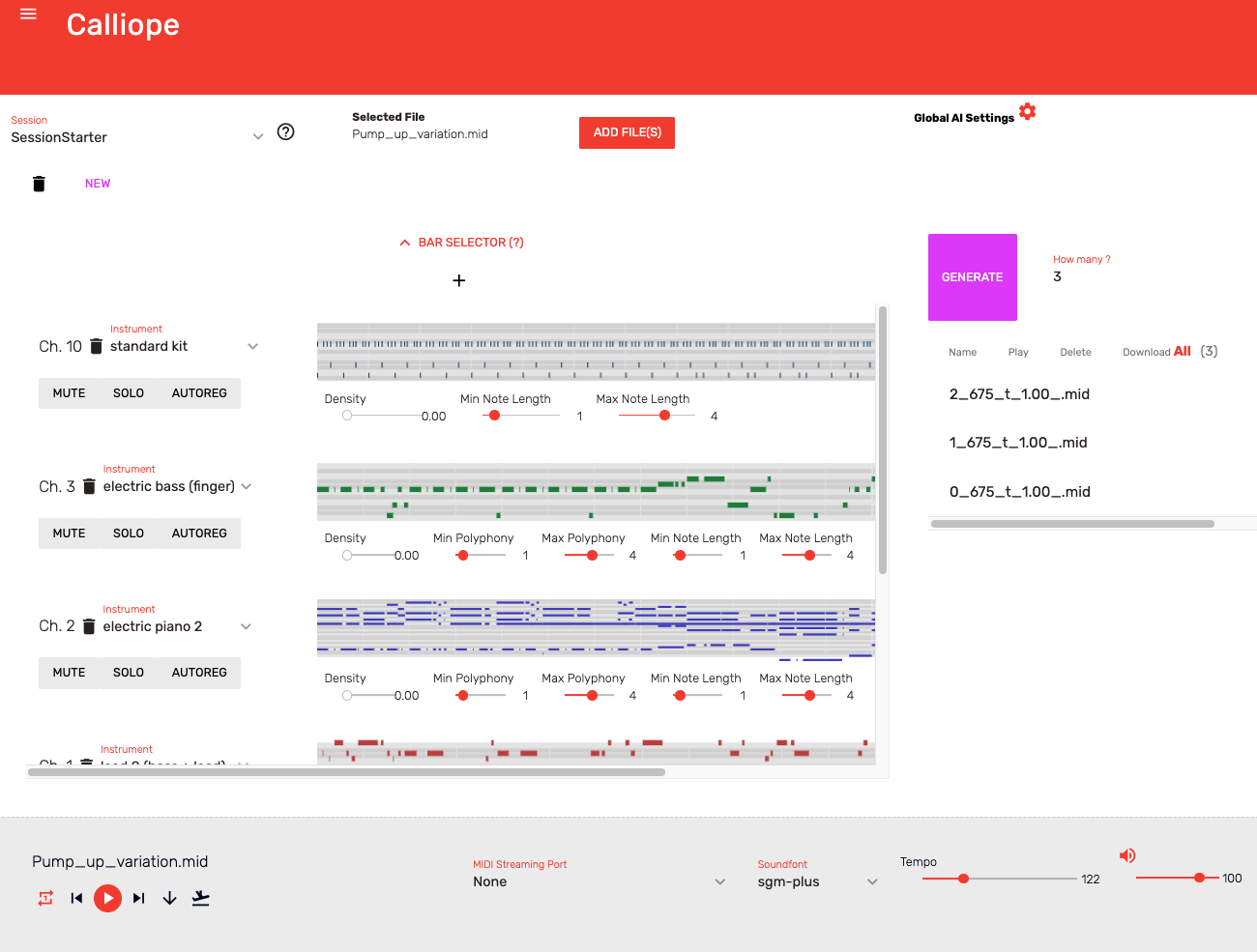}
	\caption{Calliope's Interface}
	\label{fig:maininterface}
\end{figure*}

\begin{abstract}
\begin{quote}
With the rise of artificial intelligence in recent years, there has been a rapid increase in its application towards creative domains, including music. There exist many systems built that apply machine learning approaches to the problem of computer-assisted music composition (CAC). Calliope is a web application that assists users in performing a variety of multi-track composition tasks in the symbolic domain. The user can upload (Musical Instrument Digital Interface) MIDI files, visualize and edit MIDI tracks, and generate partial (via bar in-filling) or complete multi-track content using the Multi-Track Music Machine (MMM). Generation of new MIDI excerpts can be done in batch and can be combined with active playback listening for an enhanced assisted-composition workflow. The user can export generated MIDI materials or directly stream MIDI playback from the system to their favorite Digital Audio Workstation (DAW). We present a demonstration of the system, its features, generative parameters and describe the co-creative workflows that it affords.
\end{quote}
\end{abstract}

%Computer-Assisted Composition
%Creative Music System
%Co-Creativity
%Deep Learning
%Interactive System
%Musical Metacreation
%Software System
%Web System

\section{Introduction}\label{sec:introduction}

The development of computer-assisted composition (CAC) systems is a research activity that dates back to at least the works by IRCAM on OpenMusic \cite{assayag1999computer}. CAC is a field that is concerned with developing systems that are capable of automating partially or completely the process of music composition. There exist several compositional tasks a system can address: multi-track pattern generation, multi-track complete generation, rhythm generation, harmonization, chord progression generation, melody generation, interpolation, form-filling, orchestration and interpretation.
%Each of these tasks can be realized given a conditioning or not on prior musical sequences or on supporting instrumentation.
Many machine learning-based (ML) systems have been developed for computer-assisted composition including: Flow Machines \cite{pachet2004design}, Style Machine \cite{anderson2013generative}, Magenta Studio \cite{roberts2019magenta}, Manuscore \cite{maxwell2012manuscore}, Morpheus \cite{herremans2017morpheus}; demo systems such as Sornting, DrumVAE \cite{thio2019minimal}, DeepDrum \cite{makris2018deepdrum} and commercial systems such as AIVA \footnote{https://www.aiva.ai/}, Spliqs \footnote{https://www.spliqs.com/} and Melody Sauce \footnote{https://www.evabeat.com/}. 
Magenta Studio, DrumVAE and Sornting deploy algorithms based on the MusicVAE model \cite{roberts2018hierarchical}. DeepDrum proposes an adaptive neural network model for better capturing drum rhythms. Flow Machines and Style Machine employ Markov models. Manuscore uses a cognitive architecture and Morpheus combines a tensor model with constraint rules. Finally, AIVA, Spliqs and Melody Sauce employ proprietary algorithms; the first two for generating conventional multi-track music, and the last one, for melody creation. %Magenta Studio can be used as a desktop application as well as deploy as Ableton Live plugins.
Calliope differentiates itself by using a Transformer model called the Multi-Track Music Machine (MMM). MMM, trained on half a million MIDI files \cite{ens2020mmm}, offers genre-agnostic batch-enabled generative capabilities. Its rich multi-level attribute controls combined with bar infilling enables to tackle many composition tasks at once.
%The application enable MIDI streaming to a MIDI port, including to a port on a DAW. 
Calliope has been released publicly and is being used by a variety of composers for artistic purposes and in the context of usability and acceptability evaluation studies. The project is available at    \textcolor{red}{\textbf{https://metacreation.net/calliope}}. %The web page contains a short video demonstration of the system, a list of music generated examples and a form to get onboarded to use the system.

\section{System Description}\label{sec:description}
Building on top of Apollo, our interactive web environment that makes corpus-based music algorithms usable for training and generation via a convenient graphical interface \cite{tchemeubeapollo}, Calliope (Figure \ref{fig:maininterface}) is narrowed down for MIDI manipulation in the browser, generative controllability of the MMM model, batch generation of partial or complete multi-track compositions and interoperability with other MIDI-based systems. The aim is to enable users to effectively co-create with a generative system. Calliope is built in Node.js, the Web stack (HTML, CSS, Javascript) and MongoDB. It is made interoperable with the MMM pre-trained model via a Python process runtime.

\subsection{MIDI Viewing and Playback}
\begin{figure}
	\centering
	\includegraphics[width=1\linewidth]{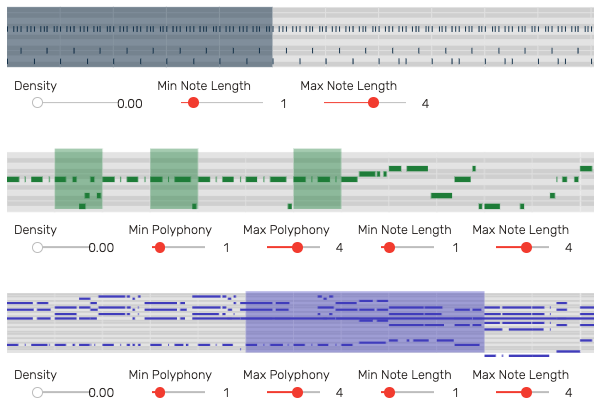}
	\caption{Multi-Track Piano Roll with Bar In-Filling}
	\label{fig:pianoroll}
\end{figure}

MIDI notes from any uploaded MIDI file can be visualized in a piano roll format (Figure \ref{fig:pianoroll}). Metadata info such as the MIDI channel number and assigned MIDI instrument can also be viewed and edited. The MIDI player supports the General MIDI (GM) standard for MIDI playback and the capacity to select from a list of soundfonts.

\subsection{Conditioned Music Generation}
Generation is achieved using the Multi-Track Music Machine \cite{ens2020mmm}. Because of its design which uses bar selection, and a set of \textit{global} and \textit{local} attribute controls, the model can accommodate a variety of compositional tasks.

\subsubsection{Bar Selection}
MMM's primary mode of generation is \textit{bar in-filling}. The model can generate note patterns for bars in a given multi-track MIDI file. A subset of bars across the multi-track content can be selected for generation by visually highlighting them (Figure \ref{fig:pianoroll}). It is also possible to temporarily edit the MIDI file by deleting or adding tracks. This is useful to perform generation on a subset of the MIDI tracks or to generate a new track for a given MIDI file. Generated music for a particular subset of bars is constrained on musical information that precedes those bars (within the given track) and on musical information found within the neighboring tracks. 

\subsubsection{Global Parameters}
MMM offers the following \textit{global} (model-level) generation parameters (Figure \ref{fig:globalparams}):

\begin{itemize}
 \item \textbf{Temperature} [0.8, 1.2]: Also called typicality, this \textit{float} value determines how much the structure of the generated MIDI content is closer (conservative) or farther (experimental) to what the MMM model is most likely to generate. Technically, it corresponds to the the temperature in the sampling of the neural network.
 \item \textbf{Polyphony Hard Limit} \{1-6\}: The global maximum number of simultaneous notes the system can generate at any given moment.
 %This value overrides the values set on the track-specific polyphony range parameters.
 \item \textbf{Percentage} \{0-100\}: This parameter controls how much of the existing MIDI content is preserved or replaced by the generation. This is done based on the number of tracks per step and bars per step. For example, for tracks per step and bars per step each 4, and percentage at 25, the model will process only 4 out of 16 bars to be generated at each generation step.
 \item \textbf{Model Dimensions} \{1-8\}: The dimension of the model in bars. This is the window size used by the model to process MIDI input data for generation. The default value is 4 corresponding to a 4-bar window.
 \item \textbf{Tracks per Step} \{1-8\}: Number of tracks being processed at each generation step. The default value is 4.
 \item \textbf{Bars per Step} \{1-8\}: The number of bars processed within each track at each generative step. The default value is 2.
 \item \textbf{Max Steps} \{0-8\}: The maximum number of generation steps. This value can be used to avoid memory overload. When it is set to zero, it is ignored by the system.
 \item \textbf{Tempo}: The resulting tempo for the generated output as a positive \textit{integer} value.
\end{itemize} 

\subsubsection{Track Parameters} In addition to model-level parameters, MMM offers a set of \textit{local} (track-specific ) music-based generation parameters (Figure \ref{fig:pianoroll}). Such parameters are available to be specified for each track of a given MIDI file. They are: 

\begin{figure}
	\centering
	\includegraphics[width=0.6\linewidth]{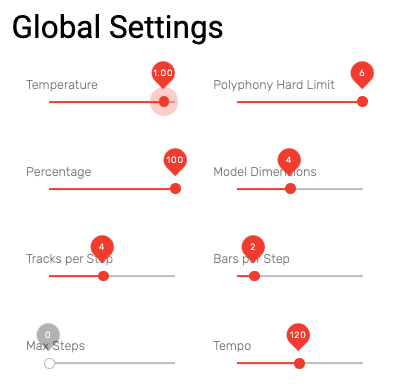}
	\caption{MMM's Global Parameters}
	\label{fig:globalparams}
\end{figure}

\begin{itemize}
    \item \textbf{Instrument Type}: The type selector is composed of a set of 128 instrument types and 8 instrument groups following the MIDI GM Standard \footnote{https://en.wikipedia.org/wiki/General\_MIDI}. It conditions MMM to generate in the style of the chosen instrument. For example, if \textit{violin} is selected, MMM generates a MIDI pattern to be played by a violin instrument. This is especially convenient to differentiate the \textit{percussion} group (e.g. drums) vs other instrument track types (e.g. \textit{guitar}, \textit{strings}, \textit{synth lead} groups).
    \item \textbf{Note Density} [0-10]: The number of notes generated per bar size. The higher this value, the more likely the model is to generate bars with a high total number of notes. A value of zero means that the note density is set at random by the model for each generation request.
    \item \textbf{Polyphony Range} \{0, 1, 2, 3, 4, 5, 6\}: the number range of simultaneous notes used by the model as a soft constraint for generation. The upper limit of this parameter is automatically overriden by the value of the "Polyphony Hard Limit" global parameter.
    \item \textbf{Note Duration Range} \{Any, 1/32, 1/16, 1/8, 1/4, 1/2, Whole\}: Note duration values are defined in accordance to the Western music notation. For example, 1/16 corresponds to a note duration equivalent to a sixteenth note.
\end{itemize} 

\subsection{Batch Generation of Music Outputs}
\begin{figure}
	\centering
	\includegraphics[width=0.75\linewidth]{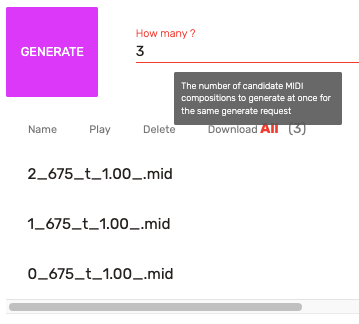}
	\caption{Batch Number for Generation}
	\label{fig:batchnranking}
\end{figure}

Batch generation of musical outputs is implemented by passing a \textit{batch\_size} parameter (Figure \ref{fig:batchnranking}) to the MMM Python interface which offers batch support natively. The ability to batch generate means that the user can quickly explore alternatives, including generating from a previously generated output, for a given set of control parameters. We have tested generation of 5, 10, 100, 500, 1000 music samples  at a time. These generations can be done within 3 seconds to 10 minutes on an average computer depending on the total note density of the music input.

\subsection{Ranking}
It is possible to rank a collection of generated MIDI files against a selected one. This is useful to informally evaluate the quality of the model generations. We employ a ranking algorithm which statistically quantifies the similarity of a generated output MIDI file against the set of other MIDI files \cite{ens2020quantifying}. This enables assessing accuracy or reliability of the MMM model for style imitation tasks. From an interaction point-of-view, it helps the user explore the variability in similarity among MIDI files and effectively apply filter operations on the set of files. This is  especially useful in the context of large set of generated files (e.g. set of 50 files and up).

\section{Co-Creative Interaction}
\begin{figure}
	\centering
	\includegraphics[width=1.01\linewidth]{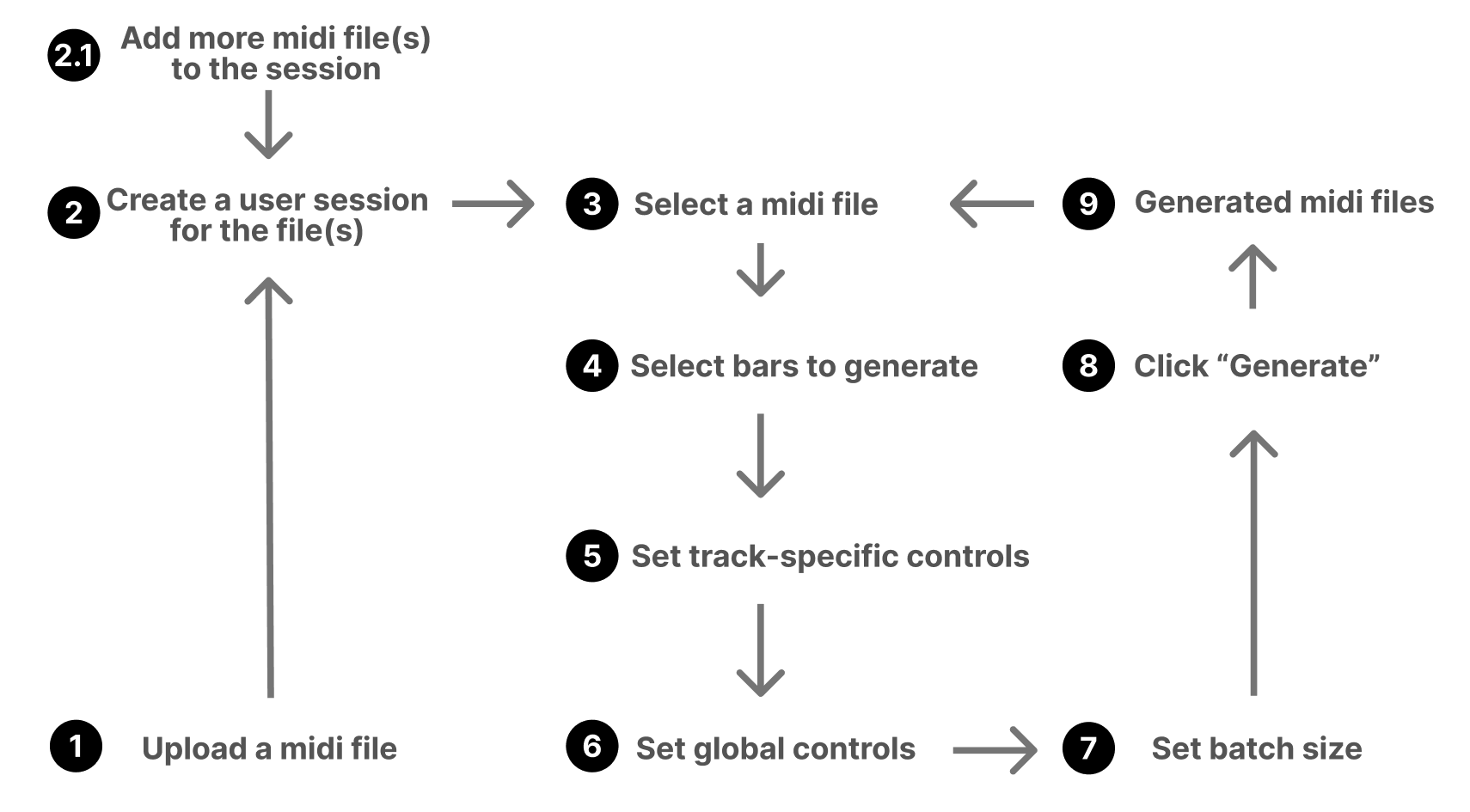}
	\caption{Compositional Workflow in Calliope}
	\label{fig:genworkflow}
\end{figure}

In terms of co-creation, the user can configure multiple attribute controls for generation (instrument type, node density, polyphony range, note length range, bar selection within a piece). Those controls set the creative context for the system to generate, allowing the user to steer the generative behavior of the model and guide the composition process. 
The system generates new musical phrases by outputting multi-track polyphonic sequences of notes for the set of selected bars and in accordance to the attribute control values. 
The user listens and analyzes the resulting output and updates the generation request accordingly. The steps involved in Calliope's interactive workflow are shown in Figure \ref{fig:genworkflow}. Generation happens within an interactive context defined by a \textit{user session} (step 2). The user session itself is defined by a seed MIDI file, which is used to kick-off the first generation. 
%Specifically, the user works in project sessions defined by a starting seed MIDI file (step 2). 
The connection from steps 9 to 3 highlights how generated outputs can themselves later be fed back into the system as seed MIDI files for new user sessions. This enables more complex workflows for the user within Calliope.

%The system enables tasks including single-track and multi-track melody generation, chord generation and rhythm generation (via drums selection), harmonization, conditioned and unconditioned generation (e.g. generating from scratch starting with an empty pattern). 

%This process is enhanced by an interactive continuous process that alternates between music generation and playback listening by the user. 

%The next diagram illustrates these interactive relationships.

\label{sec:page_size}

\subsection{MIDI Streaming} Additionally, it is possible for the user to stream MIDI playback to their favorite DAWs to assign playback to their own project session instrumentation. Calliope can be integrated with the user's digital studio (e.g. Ableton) via a MIDI port accessible in the MIDI player. This provides a unique opportunity for the user to interface their native environment with a generative system. Users can stream playback of generated MIDI files to their preferred instrumentation and sounds, including applying their existing preferred signal chains to the live output audio stream. This opens up new areas for workflow experimentation given a computer-assisted composition framework. Alternatively, they can download the MIDI files from Calliope and import them back into new or existing DAW project sessions.

\section{Conclusion}
We presented the Calliope system, a co-creative interface for multi-track music generation. We presented its features including the ability to view and play MIDI files, the ability to select bars to guide partial generation, and complete set of global (model-level) and local (track-specific) controls and how their combination allows users to tackle a broad range of compositional tasks. We situated our system with respect to other existing CAC systems and discussed the co-creative aspect of the system along with the compositional workflow it affords. The Calliope system is at the beta phase and we are working on its next version. More future work includes an ongoing evaluation study of the system along human factors including usability, user experience on feeling of trust, authorship, controllability and measured of technology acceptance among amateurs and professional composers.

\section{Author Contributions}
Bougueng R. T. was in charge of writing the manuscript and developed a significant part of the system. Ens J. developed the integration of the algorithm used by the system and assisted in making its use functional. Pasquier P. supervised the entire research process and provided direction and guidance to the project implementation. All authors participated in the writing of this manuscript and are listed in alphabetical order.

\section{Acknowledgments}

We would like to acknowledge SSHRC, NSERC and MITACS for their contribution in helping fund this research. We would also like to acknowledge Tara Jadidi as a code contributor to Calliope MIDI Viewer/Editor and bar selection feature.

%\appendix{\LaTeX{} and Word Style Files}\label{stylefiles}

%The \LaTeX{} and Word style files are available on the ICCC-13
%website, {\tt http://computationalcreativity.net/iccc2013/}.
%These style files implement the formatting instructions in this
%document.

%The \LaTeX{} files are {\tt iccc.sty} and {\tt iccc.tex}, and
%the Bib\TeX{} files are {\tt iccc.bst} and {\tt iccc.bib}. The
%\LaTeX{} style file is for version 2e of \LaTeX{}, and the Bib\TeX{}
%style file is for version 0.99c of Bib\TeX{} ({\em not} version
%0.98i).

%The Microsoft Word style file consists of a single template file, {\tt
%iccc.dot}. 

%These Microsoft Word and \LaTeX{} files contain the source of the
%present document and may serve as a formatting sample.  

\bibliographystyle{iccc}
\bibliography{iccc}

\end{document}